\documentclass{aa}
\usepackage{graphicx}  

\def\cmmt{~cm$^{-3}$}
\def\cmmd{~cm$^{-2}$}
\def\smu{s$^{-1}$}       
\def\mum{$\mu$m}
\newcommand{\gsim}{\raisebox{-.4ex}{$\stackrel{>}{\scriptstyle \sim}$}}
\newcommand{\lsim}{\raisebox{-.4ex}{$\stackrel{<}{\scriptstyle \sim}$}}
\def\HII{\ion{H}{ii}}
\def\NeII{\ion{Ne}{ii}}
\def\NeIII{\ion{Ne}{iii}} 
\def\NII{\ion{N}{ii}} 
\def\NIII{\ion{N}{iii}}   
\def\SIII{\ion{S}{iii}}   
\def\OIII{\ion{O}{iii}}  
\def\ne{$n_\mathrm{e}$}

\def\Teff{$T_\mathrm{eff}$}
\def\Av{$A_\mathrm{V}$}

\def\apj{ApJ}
\def\apjs{ApJS}
\def\aap{A$\&$A}
\def\aapr{A$\&$AR}
\def\mnras{MNRAS}

\def\aj{AJ}
\begin{document}
\title{
Large scale ionization of the Radio Arc region                    
by the Quintuplet and the Arches clusters 
\thanks{Based on observations with ISO, an ESA project 
with instruments funded 
by ESA Member States (especially the PI countries: France, Germany, the
Netherlands and the United Kingdom) and with the participation of ISAS 
and NASA.}
}

\author{ Nemesio J. Rodr\'{\i}guez-Fern\'andez 
\and     Jes\'us Mart\'{\i}n-Pintado 
\and     Pablo de Vicente  
}

\offprints{N.J.~Rodr\'{\i}guez-Fern\'andez, \email{nemesio@oan.es}}

\institute{Observatorio Astron\'omico Nacional, Instituto Geogr\'afico
Nacional, Apdo. 1143, E28800 Alcal\'a de Henares, Madrid, Spain
}

\date{Received /Accepted}

\abstract{We present an analysis  of
selected fine structure lines 
(\NeII~12.8\,\mum, \NeIII~15.6\,\mum, \SIII~18.7 and 33.5\,\mum,
\OIII~52 and 88\,\mum, \NII~122~\mum~  and \NIII~57\,\mum) 
observed  with the {\it Infrared Space Observatory} (ISO)  toward the
Radio Arc in the Galactic center region (GCR).
Most of the data    were retrieved from the ISO Data  Archive.
We study the density of the ionized gas and the 
large scale  ionization structure in a region
of $\sim 30\times30$ pc$^2$ by means of the
\OIII~52\,\mum/88\,\mum, \SIII~18.7\,\mum/33.5\,\mum,
\NeIII~15.6\,\mum/\NeII\,12.8\,\mum~ and
\NIII~57\,\mum/\NII~122\,\mum~ line ratios. 
The electron densities (\ne) derived from the 
\OIII~ lines ratio indicate the presence of  diffuse  ionized material
with \ne~  of $\sim 10^{1.8-2.6}$~\cmmt.
The \NeIII~15.6\,\mum/\NeII~12.8\,\mum~ and 
\NIII~57\,\mum/\NII~122\,\mum~ line ratios
vary from source to source
from 0.05 to 0.30 and from 0.3 to 2.5, respectively.
The \NIII~57/\NII~122~\mum~ ratio show two clear gradients,
one pointing toward the Quintuplet cluster and the other pointing
toward the Arches cluster.
We have used a simple model to explain the ionization structure
observed in the \NIII/\NII~ and \NeIII/\NeII\ lines ratios.
The model shows that  the large scale  ionization
of the whole region can be accounted for
by the UV radiation produced by the Quintuplet and the Arches cluster.
Any other ionization mechanism should play a minor      role. 
We also  investigate the influence of the clusters on the
bubble of warm dust (hereafter Radio Arc Bubble, RAB)
seen in  the {\it Midcourse Space Experiment} 
(MSX) infrared images.
We  find that the warm dust is well  correlated
with the ionized gas indicating that the dust is also heated by
the radiation from both clusters.
Furthermore,  the elliptical rather than circular symmetry of some
structures like the Thermal Filaments 
can also be explained when one considers the combined effects of
both the Arches and the Quintuplet clusters.
We have also found that the RAB  
is filled with continuum emission of hard X rays and with
emission from  the  6.4 keV line of neutral or low ionized Fe.
We briefly discuss  the implications of these findings
on the structure and morphology of the GCR interstellar medium
and the possible origin of the RAB.
\keywords{Galaxy: center --  ISM: \HII~ regions -- 
ISM: dust,extinction -- ISM: individual objects: Radio Arc --
Infrared: ISM: lines and bands -- X-rays: ISM: lines and bands}
}

\titlerunning{Ionization of the Radio Arc}
\authorrunning{N.J. Rodr\'{\i}guez-Fern\'andez et al.}
\maketitle

\section{Introduction}
With the aim of understanding the thermal balance and ionization
of the interstellar medium (ISM)
 in the Galactic center region (GCR),
we have undertaken a systematic study
at millimeter and infrared wavelengths of a sample of clouds
distributed along the Nuclear Bulge (in notation of Mezger et al. \cite{Mezger96}) 
and the ``Clump 2" complex.
These clouds  exhibit large column densities of warm H$_2$ (with temperatures
of $\sim 150$\,K) which   constitute an important             
fraction of the total column densities of molecular gas
(Rodr\'{\i}guez-Fern\'andez et al. \cite{RF01}).
In addition, the comparison of fine-structure lines of several ionized atoms 
with  hydrogen radio recombination lines indicate that
the GCR ISM should be rather inhomogeneous and that the 
the ionized gas could be distributed, for instance, in relatively low
density ionized  bubbles (Mart\'{\i}n-Pintado et al. \cite{MP00a};
Rodr\'{\i}guez-Fern\'andez et al. in prep.).

The two    sources of our sample 
with more intense fine structure lines (\object{M+0.16--0.10} and 
\object{M+0.21--0.12}, see Mart\'{\i}n-Pintado et al. 2000a) 
are  located in the vicinity of the  \object{Radio Arc},
which is one of  the most  prominent radiocontinuum
features in the GCR. The Radio Arc, located  
at $l\sim 0\fdg17$, is  perpendicular to the Galactic plane and it is 
apparently connected to Sgr A by a ``bridge" of
radiocontinuum emission
at positive galactic latitude (see e.g. G\"usten \cite{Gusten89}).
VLA  observations showed   that the Radio Arc
is  composed    by   long  and  thin straight  filaments
that emit non-thermal radiation (Yusef-Zadeh, Morris \& Chance \cite{YZ84}; 
Yusef-Zadeh, Morris \& Gorkom \cite{YZ87}).
These filaments are  usually known as the 
\object{Non-Thermal Filaments} (NTFs) and indicate the presence  of 
a strong poloidal component of the magnetic field in the GCR
(Yusef-Zadeh, Morris \& Chance \cite{YZ84}; Yusef-Zadeh \& Morris 
\cite{YZ87a}, \cite{YZ87b}).

On the other hand, the ``bridge" is composed  by a set of  arched filaments
that emit  thermal radiation (\object{Thermal or Arched Filaments}).
In the vicinity of the NTFs and probably interacting with them
there  are other prominent  features that also  emit thermal   radiocontinuum.
These are \object{G0.18--0.04} (the Sickle), which is  approximately located
where the NTFs cross the galactic plane, and a gun-shaped
source  located south of the Sickle known as the Pistol Nebula
(\object{G0.15--0.05}).
The thermal nature of the Sickle, the Pistol Nebula and the Thermal filaments
has been established by the association  of the radiocontinuum emission
with radio recombination line emission (Pauls et al. \cite{Pauls76}; 
Paul \& Mezger \cite{Pauls80}; Yusef-Zadeh, Morris \& van Gorkom \cite{YZ87}).

In the recent years, the  origin of the ionization of these  thermal features
has been  a subject of great interest.
It was first thought that the Sickle and the Thermal Filaments
were  the surfaces of 
molecular clouds ionized by collisions between the neutrals and the
ions that spiral around the magnetic field lines associated with
the NTFs (Yusef-Zadeh \& Morris \cite{YZ87c}; Morris \& Yusef-Zadeh \cite{Morris89};
Serabyn \& G\"usten \cite{Serabyn91}; Serabyn \& Morris \cite{Serabyn94}).
However, with the discovery of the outstanding stellar cluster  known as
the \object{``Quintuplet" or AGFL\,2004} (Nagata et al. \cite{Nagata90};
Okuda et al. \cite{Okuda90};
Glass et al. \cite{Glass90}) the effect of ultra-violet (UV) radiation
on the ionization of the Sickle was revised.
Nowadays, it is commonly assumed that this cluster is the main
ionization source of the Sickle \HII~ region (Timmermann et al. \cite{Timmermann96};
Simpson et al. \cite{Simpson97}).
Furthermore, there is increasing evidence that the Pistol Nebula is also
ionized by the Quintuplet instead of by the Pistol  star, located
in the center of
curvature of the nebula (Moneti et al. \cite{Moneti99}; Figer et al. \cite{Figer99}).

On the other hand, the \object{Object\,17} in the infrared 
survey undertaken  by Nagata et al. (\cite{Nagata93}) was found to be a cluster
of young stars (Nagata et al. \cite{Nagata95}; Cotera et al. \cite{Cotera96}). 
This cluster, also known as \object{G0.121+0.017} or the Arches cluster
have enough OB stars  to be the  ionization source of the
Thermal Filaments (Cotera et al. \cite{Cotera96}; Serabyn et al. \cite{Serabyn98}). 
However, to explain the homogeneity in the excitation observed in the fine
structure line emission, Colgan et al. (\cite{Colgan96}) proposed that more  hot stars
should be distributed rather uniformly through the region
of the Arched Filaments.
Nonetheless, the recent  analysis of fine structure
lines observed by ISO (Cotera et al. \cite{Cotera00a}) and the
recombination lines VLA  observations of Lang et al. (\cite{Lang01}), 
suggest that the Arches cluster alone could  account for the
ionization of the Arched Filaments.

Recently, mid-infrared images (7-25 \mum)  of the GCR  taken by the  
{\it Midcourse Space Experiment} (MSX)  satellite have shown
a new  morphological feature in the Radio Arc region (Egan et al. \cite{Egan98} ).
It is a ring of warm dust which is also present,
although it is rather weak, in the 20 cm  radio image
of Yusef-Zadeh \& Morris (\cite{YZ87c}).
Although it is not yet clear 
its actual morphology (bubble, ring or cylinder),
in the following we will call it the  Radio Arc Bubble (RAB).
The RAB is also seen in the IRAS HIRES maps
(see Simpson et al. \cite{Simpson97}) where it appears as a ring of 
warm dust emitting mainly  at  25 and 60 \mum.
The PAHs emission detected by ISO in the RAB shows
the presence of  carbon-rich material in the bubble (Levine et al. \cite{Levine99}).
The part of the Sickle which is  perpendicular to the NTFs
just lie in  the north-west edge of the \object{RAB},
suggesting that the Sickle 
is part of this large structure of warm  dust.
The geometrical  center of the RAB is located just on the most intense
NTF at $(l,b) \sim (0\fdg16, -0\fdg11)$,
north-west of the non-thermal maximum  (\object{G0.16-0.15}).

Figer et al. (\cite{Figer99}) suggested that winds and radiation pressure
from O-type stars in the Quintuplet cluster
or even a supernova explosion cloud have blown the dust bubble.
They proposed that the offset of the Quintuplet
cluster from the center of the RAB could be due    to the
relative motion of the stellar cluster with respect to the ISM or 
to       large-scale density fluctuations in the medium where 
the bubble is expanding.
However,
the nearly circular shape of the RAB   seen in the radio
images of Yusef-Zadeh \& Morris (\cite{YZ87c}) rules out the 
hypothesis of a density gradient.

The  purpose of this paper 
is to test the scenario   of a very inhomogeneous medium
---proposed by Mart\'{\i}n-Pintado et al. (\cite{MP00a}) 
to explain the ionization of \object{M+0.12--0.12} 
and \object{M+0.16--0.10}---
by  studying the large scale  ionization structure
in the region of the Radio Arc.
We have   analyzed selected  fine structure lines of the
observations taken with the
{\it Infrared Space Observatory} (ISO; Kessler et al. \cite{Kessler96}) towards
the Radio Arc.
Previous  works were 
restricted to the Sickle and the Pistol Nebula.
We study the Sickle  in the broader context
of the RAB and  investigate  the possible connection
between the ionization in the 
Radio Arc area and  the  existence of the RAB.        
The paper is organized as follows:
Section 2 defines the sample of sources selected from the ISO
Data Archive (IDA) and Sect. 3 comments on the data reduction and results.
The analysis of the results is  presented in Sect. 4 and,  
 with the help of photoionization models, in Sect. 5.
The results are discussed in Sect. 6 and 7.
Finally, Sect. 8 summarizes the main  conclusions.


\section{Observational  data}

\begin{figure*}[]
%
%
%
\vspace{8cm}
\caption{Position of the sources observed by ISO SWS and LWS
overlaid on the band E (18.2-25.1 \mum) MSX image.
The display is logarithmic from 4 $10^{-6}$ to 8 $10^{-4}$ 
W\,m$^{-2}$\,sr$^{-1}$.
The circle delineates the edges    of the warm dust  bubble  (RAB,
see also Fig.\ref{fig_UX}).
The strong features to the north-west of the RAB are
the Sickle and the Pistol Nebula}
\label{fig_obs}
\end{figure*}

To study the large scale ionization structure 
of  the Radio Arc region and its possible
connection to the ring or bubble of warm dust (hereafter RAB)  we
have used {\it Short Wavelength Spectrometer} (SWS; de Graauw et al. \cite{deGraauw96})
and {\it Long  Wavelength Spectrometer} (LWS; Clegg et al. \cite{Clegg96})
ISO data of our sources M+0.16-0.10 and M+0.21-0.12.
We have also retrieved other LWS and SWS observations towards  this region
from the ISO Data Archive (IDA).
We have  analyzed  the  \NeIII~15.6\,\mum, \NeII~12.8\,\mum~
and the \SIII~ 33.5 and 18.7 \mum~ fine structure lines observed with
the SWS.
In the LWS observations we have analyzed the 
fine structure  lines of \NIII~  at 57\,\mum, \NII~  at 122\,\mum~
and \OIII~ at 52 and 88\,\mum.
Table~\ref{tab_obs} lists IDA identification numbers (TDTs),  
coordinates and  observing modes (AOTs), for all the selected sources.     
Figure~\ref{fig_obs} shows the position of the data points  overlaid on
the 18.2-25.1~\mum~ band  MSX image in which  the RAB is clearly
visible.

Positions M1 (\object{M+0.16-0.10})  and M2 (\object{M+0.21-0.12}) 
are the two clouds from  our sample located in the 
Radio Arc region (Mart\'{\i}n-Pintado et al. \cite{MP00a};
Rodr\'{\i}guez-Fern\'andez et al. \cite{RF01}).
M1 lies close to the geometrical center of the RAB over the most intense  NTF. 
Positions B1 to B3 lie in the southern part of the \object{RAB}
(see Levine et al. 1999)
while B4 and B5 are located in the northeast edge of the RAB, close to M2.
Q1 and Q2 are the  Quintuplet sources \object{GCS3 II}  and 
\object{GCS4} (see Chiar et al. \cite{Chiar00}), S1  and S2 are two
observations toward the Sickle and P toward the Pistol Nebula.
R is a raster map of 4$\times$9 points (4 rows, 9 columns),
with a spacing of 50$^{''}$ and 
rotated 50\degr~ north from    west.  
In the following, individual points in the raster
will be named as  R$xy$, i.e., R11, R12,...

\begin{table}[]
 \caption[]{Observations summary}
 \label{tab_obs}
 \begin{tabular}{lllll}
   \hline
   \noalign{\smallskip}
   Position & RA$^{\mathrm{a}}$  & Dec$^{\mathrm{a}}$ &
                           TDT$^{\mathrm{b}}$  &  AOT$^{\mathrm{c}}$\\
   \noalign{\smallskip}
   \hline
   \noalign{\smallskip}
     M1         & 17:46:24.9& -28:51:00.0& 48502207& SWS    02 \\
                &           &            & 48502309& LWS    01 \\
     M2         & 17:46:34.9& -28:49:00.0& 48502612& SWS    02 \\
                &           &            & 49401613& LWS    01 \\
     R $^{\mathrm{d}}$   & 17:46:09.4& -28:48:07.4& 67700702& LWS 01 \\
     B1         & 17:46:11.0& -28:54:36.3& 46301403& SWS 01 \\
     B2         & 17:46:43.8& -28:52:53.9& 46300901& SWS 01 \\
     B3         & 17:46:31.4& -28:55:48.8& 49800804& SWS 01 \\
     B4         & 17:46:32.8& -28:47:42.2& 69601107& LWS   01 \\
     B5         & 17:46:42.5& -28:49:01.3& 69601107& LWS   01 \\
     Q1         & 17:46:15.7& -28:49:47.0& 29702147& SWS 01 \\
     Q2         & 17:46:14.8& -28:49:34.0& 28701246& SWS 01 \\
     S1         & 17:46:14.2& -28:47:47.1& 67700503& SWS 01 \\
     S2         & 17:46:15.4& -28:48:07.0& 46400904& SWS 02 \\
     P          & 17:46:15.2& -28:50:04.0& 84101302& SWS  01 \\
   \noalign{\smallskip}
   \hline
 \end{tabular}
\begin{list}{}{}
\item[$^{\mathrm{a}}$] J2000 coordinates
\item[$^{\mathrm{b}}$] Observation number
\item[$^{\mathrm{c}}$] Astronomical Observing Template (instrument and
 observing mode)
\item[$^{\mathrm{d}}$]  4 $\times$ 9 raster map. The coordinates are
the center of the raster.
\end{list}
\end{table}
%


\section{Data reduction and results}

The observations consisted of SWS full grating spectra (SWS01),
SWS grating spectra of selected lines (SWS02),
and LWS full grating spectra (LWS01).
The telescope aperture at the wavelengths of the lines of interest are
listed in Table~\ref{tab_sws} and Table~\ref{tab_lws}.
M1 and M2  have been processed through the  Off Line 
Processing (OLP) version  7.0.
The other observations have been automatically reprocessed when retrieval
from the IDA  with OLP versions 9.1 to 9.5.
Further reduction and analysis have been carried out  with ISAP 2.0.

\subsection{SWS AOTs 01 and 02}

As a first stage, we have removed bad data points. 
Afterwards, for the AOT 01 observations,
we have shifted the different detectors
to a common level and we have averaged across scan directions and
detectors.  For the AOT 02 observations we have averaged across 
scan directions before shifting the different detectors
to a common level and averaging across detectors.
Baselines of order 1 were removed from the spectra.
The fluxes of the lines with their corresponding rms errors
as derived from Gaussian fits  are listed in Table~\ref{tab_sws}.
Errors in the  line fluxes due to calibration uncertainties are 
smaller than 25$\%$ for the \NeIII~ and \NeII~lines and less
than 20$\%$ and 30$\%$ for the \SIII~ 18.7 and 33.4 \mum, respectively
(Leech et al. 2001).

\begin{table}[]
 \caption[]{Fluxes derived from Gaussian fits to the lines
      observed with the SWS in units of 10$^{-20}$ W \cmmd.
      Numbers in parentheses are the rms
      errors of the last significant digit.}
 \label{tab_sws}
 \begin{tabular}{lllll}
  \hline
  \noalign{\smallskip}
           & \NeII  & \NeIII & \SIII & \SIII \\
  $\lambda$(\mum) & 12.81  & 15.55  & 18.71  & 33.48  \\
  Beam (arcsec$^2$)  & $14 \times 27$ & $14 \times 27$ &
                                     $14 \times 27$ & 20$\times$ 33 \\
  \noalign{\smallskip}
  \hline
  \noalign{\smallskip}
  M1          &  41(2) & 9.3 (13) & 14.4(10) & 102(3)  \\
  M2          & 139(2) & 20(2)    & 63(2)    & 425(12) \\
  B1                   & 92(8)  & 19(2)    & 41(2)    & 236(8)  \\
  B2                   & 179(5) & 30(4)    & 80(4)    & 640(30) \\
  B3                   & 138(12)& 10(4)$^{\mathrm{a}}$    & 56(3)  & 480(20) \\
  Q1                   & 439(12)& 149(6)   & 270(6)   & 600(20) \\
  Q2                   & 122(12)& 35(12)   & 267(6)   & 373(13) \\
  S1                   & 610(30)& 125(11)  & 340(20)  &1860(90) \\
  S2                   & 491(5) & 90(2)    &  ---     & ---     \\
  P               & 500(29)& 227(6)   & 319(8)   &879(14)\\
  \noalign{\smallskip}
  \hline
 \end{tabular}
 \begin{list}{}{}
  \item[$^{\mathrm{a}}$] Low signal-to-noise ratio ($\sim 3$)
 \end{list}
\end{table}

\subsection{LWS AOT 01}

For all the LWS01 observations, detector SW2 showed  memory effects at
wavelengths shorter than 55 \mum. 
To analyze the \OIII\,52 \mum~ line we have
treated the two scan directions independently, averaging only  across scans.
For both scan directions the fluxes are in agreement within 15 $\%$.
The fluxes of the \OIII~ 52 \mum~ lines  listed in Table~\ref{tab_lws}
are the average of the fluxes derived independently for the two
scans directions.
For the lines at other wavelengths we have 
shifted the different scans to a common
level before averaging across the  two scan directions and across scans.
We have subtracted from the spectra  baselines of order 1. 
The line fluxes and rms errors as derived from the Gaussian fits 
are listed in Table~\ref{tab_lws}.

The \OIII\,88~\mum~ line is in the overlap region of detectors SW5 and LW1.
We have analyzed both detectors separately and  the
measured  fluxes agree within 15$\%$.
The fluxes listed in Table~\ref{tab_lws} are the average
values of both determinations.
Before fitting Gaussian profiles to the \OIII\,88~\mum~ (detector LW1)
and the  \NII\,122~\mum~ (detector LW3) lines we have defringed the spectra. 
The line flux calibration uncertainties  are expected
to be smaller  than 30$\%$ (Swinyard et al. 1996).

\begin{figure*}[]
%
%
\resizebox{12cm}{!}{\includegraphics{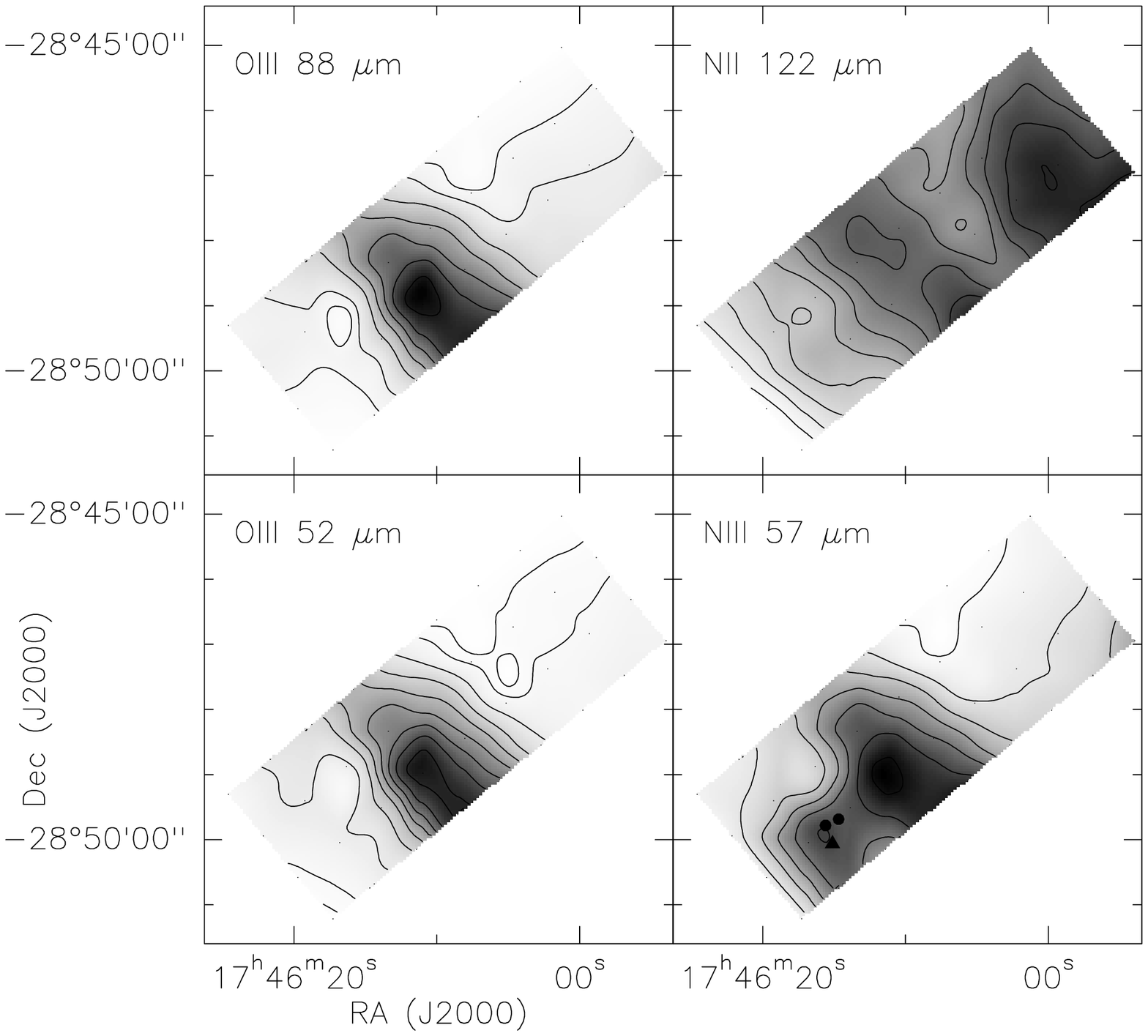}}
\caption{LWS raster maps.
{\it Upper-left:} \OIII~ 88\,\mum~  map (contour levels in units
of 10$^{-18}$~W\,\cmmd: 6, 10, and from 12.5 to  42.5 by steps of 6).
{\it Lower-left:} \OIII~ 52\,\mum~  map (contour levels in units
of 10$^{-18}$~W\,\cmmd: 7, 9, and from 14 to  50 by steps of 6).
{\it Upper-right:} \NII~ 122\,\mum~  map (contour levels in units
of 10$^{-18}$~W\,\cmmd: from 4 to 15 by steps of 1).
{\it Lower-right:} \NIII~ 57\,\mum~  map (contour levels in units
of 10$^{-18}$~W\,\cmmd: from 3.8 to  21.8 by steps of 3).
The strong elongated feature corresponds to the Sickle for all
but the \NII~122~\mum~ map.
The filled circles shown in the \NIII~57~\mum~ map indicate the location of
the Quintuplet sources GCS 4 and GCS  3II and the triangle indicate
the position of the Pistol Nebula.}
\label{fig_raster}
\end{figure*}

The raster map, R, has been reduced using the same procedure described
above and 
Fig.\,\ref{fig_raster} shows the emission maps for all the lines.
The maximum of the emission  is found towards  the Sickle for all  but the
\NII\,122  \mum~  line, which peaks in the western part of the map where the
intensity of the \NIII~ line also increases. 
The emission, however, does not completely follow the Sickle.
The part of the Sickle parallel to the NTFs that points toward the center
of the RAB is not clearly seen in any of the emission  maps.
The  \NIII~ map  also shows   a local maximum at the
position of the  Pistol Nebula.

\begin{table*}[]
 \caption[]{Fluxes derived from Gaussian fits to the lines observed
            with the LWS.
            Fluxes in units of 10$^{-19}$ W \cmmd.
            Numbers in parentheses are the rms errors of the
            last significant digit.}
 \label{tab_lws}
 \begin{tabular}{lllll}
  \hline
  \noalign{\smallskip}
      & \OIII & \NIII & \OIII  & \NII \\
  $\lambda$(\mum)   & 51.81  & 57.33  & 88.36  & 121.90  \\
  Beam (arcsec$^2$)   & 80 $\times$77 & 80$\times$77& 84$\times$76 & 78$\times$7
5  \\
  \noalign{\smallskip}
  \hline
  \noalign{\smallskip}
  M1          & 50(3) & 23(4) & 65(5) & 26(2) \\
  M2          &108(5) & 48(3) & 118(4)& 74(5) \\
  B4          &124(10)&59(6) &147(8) & 92(6) \\
  B5          & 71(8) & 38(6) &102(10) &90(6) \\
  R   & 52(7)--518(20)$^\mathrm{a}$ & 29(3)--223(10)$^\mathrm{a}$
               & 67(6)--441(9)$^\mathrm{a}$ & 40(3)--140(4)$^\mathrm{a}$\\
  \noalign{\smallskip}
  \hline
 \end{tabular}
 \begin{list}{}{}
   \item  $^\mathrm{a}$ Maximum and minimum values in the raster.
          See also Fig.~\ref{fig_raster}.
  \end{list}
\end{table*}

\subsection{Comparison with previous observations}

Some observations        presented in this paper have also  been 
previously carried out  with  other instruments.
One can compare the LWS  \OIII~52 and 88\,\mum~
and  \NIII~57\,\mum~ lines 
fluxes in the Sickle region with those measured 
by Simpson et al. (\cite{Simpson97}).
Their  points P2, P5, P3, P4, P8 and P6 are included in the LWS raster map.
Assuming extended homogeneous emission and taking into account 
the differences in the beams sizes of   both
observations, one finds that
for the sources located in the Sickle (P2, P5, P3 and P4)
the fluxes measured by Simpson et al. are
60$\%$ higher than those obtained with the LWS.
On the other hand, for P8 and P6 they are rather similar. 
These results indicate that the emission is indeed quite
homogeneous in the area of their P8 and P6 positions 
(approximately corresponding to R34 and R28 respectively)
but not in the region of    the Sickle.
However, even in the Sickle region, 
the clumpiness of the fine structure line emission in the LWS beam
is relatively small and the approximation of extended and homogeneous
emission is rather good (see also Mart\'{\i}n-Pintado et al. 2000a).
A similar  conclusion is derived 
if one compares our \OIII\,88~\mum~ and \NIII\,57~\mum~
data  with the observations of the Sickle by
Timmermann et al. (\cite{Timmermann96}). 
Taking into account that
they had  a beam of 22$^{''}$ and assuming homogeneous emission,
the expected flux in the LWS beam ($\sim 80^{''}$) is larger than that observed 
for some points and lower  for other points being on average 17 $\%$ 
larger than those measured with the LWS.
Given the different beam sizes and
the flux calibration uncertainties, we
conclude that the LWS data are in good agreement 
with previous data and that the fine structure lines observed with the LWS
show a relatively uniform   emission.


\section{Analysis}

\subsection{Extinction correction}

Before undertaking further analysis of the data
it is necessary to correct them  for the dust extinction effects.
For the typical  average visual  extinction of  $\sim 25$ mag
measured towards the GCR 
(Catchpole et al. 1990;  Schultheis et al. 1999) even at mid and far IR
wavelengths the extinction corrections are non negligible.
We can use different methods to estimate the  extinction affecting  our data.
One possibility is to use the ratio of the \SIII~
lines at 18.7 and 33.5\,\mum~
to derive lower limits to the extinction
for  the sources where these two lines are available.
This is because the \SIII\,18.7/33.5   ratio cannot  be  lower
than the theoretical  limit of 0.5  for the case of very low density. 
If the measured ratio is lower than the predicted lower limit, 
this indicates a higher   extinction  for the \SIII~18.7 \mum~  
than for the \SIII~  33.5 \mum~ line.
From our \SIII~ data, we  derived a lower limit to the visual extinction
assuming extended emission for all the positions of $\sim 20$  mag at most.

A direct determination for the overall extinction in the area of the Sickle 
has been recently  made     
by Cotera et al. (\cite{Cotera00b})  both by analyzing  
the stellar content and by  comparing
the radiocontinuum emission with  the  Br$\gamma$ emission.
They found an extinction at             2.2\,\mum~ of  $\sim 2.5-3.2$ mag. 
These values imply a visual  extinction of $\sim 25-32$ mag, 
in agreement with   the   large scale
studies of Catchpole et al. and Schulteis et al. and consistent 
with our lower limits.
In the following we will assume that our data is affected by an average 
visual extinction  of 30 mag.
To extrapolate the  visual extinction to the wavelengths of the 
SWS lines   we have 
used the extinction law derived by Lutz (\cite{Lutz99}) towards Sgr A$^*$.
This extinction law is almost equivalent to that of Draine (\cite{Draine89})
at the wavelengths of the lines presented in this paper.
To derive the extinction for the 
LWS lines  we have used the following  extinction law: 

\begin{equation}
\mathrm{A}_\lambda =0.014 A_\mathrm{V} (30/\lambda)^{1.5}
\end{equation}
where $\lambda$  is the wavelength in microns,
$\mathrm{A}_\lambda$ is the extinction at that wavelength, 
 and \Av~ is the visual extinction.

\subsection{Electron densities}

\begin{figure}[]
\centerline{\resizebox{7cm}{!}{\includegraphics{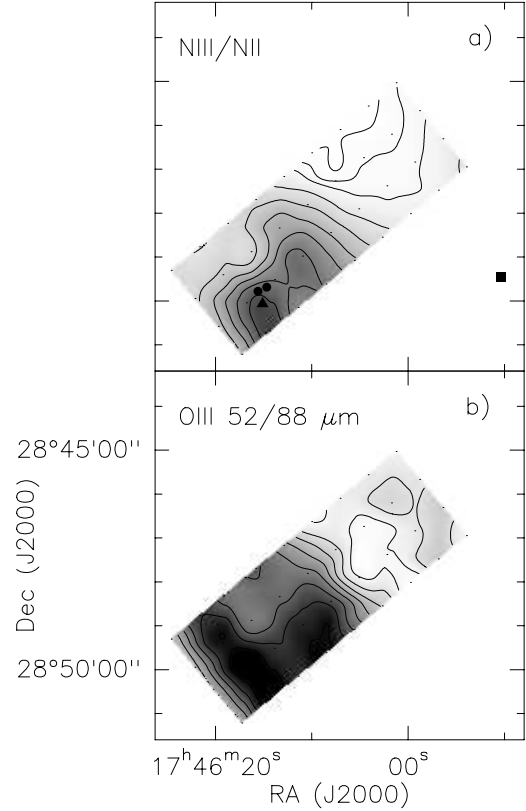}}}
\caption{ {\bf a)} \NIII~ 57 \mum~ to \NII~ 122 \mum~ ratio map
after correcting for  30 mag of visual extinction
(contour levels: 0.35, 0.45, and from 0.55 to  2.35 by 0.3).
The circles indicate the position of the Quintuplet sources
GCS 4 and GCS 3II and the triangle the location of the Pistol
Nebula. The filled square
indicates the position of the Arches cluster.
{\bf b)} \OIII~ 52 to 88 \mum~ line ratio corrected by 30 mag of visual
extinction. Contour levels: from 0.82 to  1.38 by steps of 0.08}
\label{fig_raster_coc}
\end{figure}

Table \ref{tab_ratios} lists the extinction corrected 
\OIII\,52 to 88 \mum~  and  \SIII\,18.7 to 33.5~\mum~
line intensity ratios 
(hereafter \OIII\,52/88  and \SIII\,18/33 ratios, respectively) and 
Fig.\,\ref{fig_raster_coc}b shows the map of the    extinction corrected
\OIII\,52/88 ratio  for  the raster.
The most prominent feature in this  map is  a ridge 
 that goes  northeast-southwest at the position of the Pistol Nebula.
The \OIII\,52/88 \mum~ ratio in the map ranges between 0.78 and 1.43.
Table\,\ref{tab_ratios} also list the electron densities (\ne)
derived from these ratios assuming that the lines are excited by 
collisions with electrons and that the sources are extended 
(see Rubin et al. \cite{Rubin94}).
The derived  \OIII\,52/88 ratio  implies  \ne~ between 
10$^{1.8}$  and 10$^{2.6}$ \cmmt.
From the \SIII\,18/33 ratio we can only derive an   upper limit to \ne~ of 
$\sim$100 \cmmt~ for
most  sources. For  the Pistol Nebula  and the Quintuplet sources
(P, Q1 and Q2) the derived densities
from the \SIII~ lines are as high as $10^{3.2-3.8}$ \cmmt.
These densities are higher than those derived for  
the same region from the \OIII~
lines (with a larger beam), suggesting the presence of small scale
structures with high density  in the Sickle and the Pistol Nebula
as previously mentioned in Sect. 3.3.

\begin{table*}[]
 \centering
 \caption[]{Line ratios after correcting for 30 mag. of visual extinction.
         Numbers in parentheses are the rms errors of the last
         significant digit as derived from the errors in the Gaussian fits
         to the lines.
         To calculate line ratios we have taken into account the different
         beam sizes and considered extended emission.}
 \label{tab_ratios}
 \begin{tabular}{lllllll}
  \hline
  \noalign{\smallskip}
   & \SIII   & \ne(\SIII)& \NeIII/\NeII& \OIII & \ne(\OIII) & \NIII/\NII \\
   &R(18/33) & log(\cmmt) &            & R(52/88) &log(\cmmt) &            \\
  \noalign{\smallskip}
  \hline
  \noalign{\smallskip}
  M1           &0.66(2) & $<2$  &0.148(7)  & 0.88(8) & 1.96  & 0.9(2)  \\
  M2           &0.697(13)& $<2$ & 0.094(3) & 1.04(5) & 2.17  & 0.68(6) \\
  B1           &0.82(2)  &  2.48    & 0.134(6) & ---     & ---   & ---   \\
  B2           &0.59(2) & $<2$  & 0.109(5) & ---     & ---   & ---   \\
  B3           & 0.55(2)& $<2$  & 0.047(7) & ---     & ---   & ---   \\
  Q1           & 2.12(4)& 3.3    & 0.221(3) & ---     & ---   & ---   \\
  Q2           & 3.36(7)& 3.78    & 0.19(2)  & ---     & ---   & ---  \\
  S1           & 0.86(3)&  2.70    & 0.133(4) & ---     & ---   & ---   \\
  S2           & ---    & ---     & 0.119(2) & ---     & ---   & ---   \\
  P            & 1.71(2)& 3.18   & 0.296(4) & ---     & ---   & ---  \\
  B4           & ---   & --- &--- & 0.96(8) & 2.07  & 0.68(7)  \\
  B5           & ---   & --- & ---& 0.79(10)& 1.84  & 0.45(7)  \\
  R  & ---   & --- & ---& 0.78--1.43$^\mathrm{a}$ &
                               1.81--2.56 & 0.31--2.52$^\mathrm{a}$ \\
  \noalign{\smallskip}
  \hline
 \end{tabular}
 \begin{list}{}{}
  \item  $^\mathrm{a}$ Maximum and minimum values.
                                 Typical errors lower than 10$\%$
 \end{list}
\end{table*}

\subsection{Ionization structure}
\label{sec_ion_struc}

\begin{figure*}[]
%
%
\resizebox{12cm}{!}{\rotatebox{-90}{\includegraphics{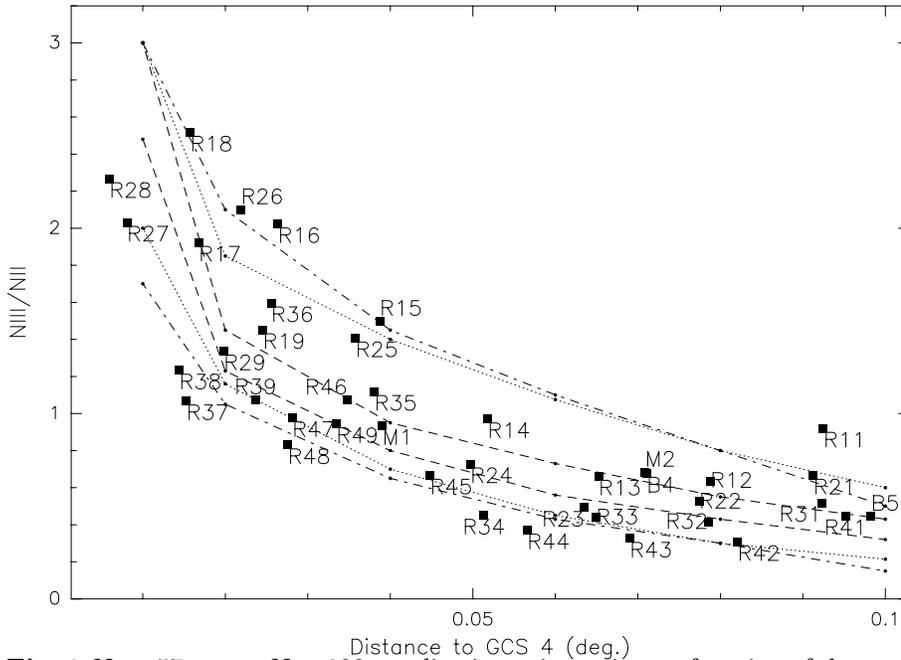}}}
\caption{\NIII~ 57 \mum~ to \NII~ 122 \mum~ line intensity
ratio as a function of  the projected distance to
the Quintuplet cluster source GCS 4 and comparison with models:
Dot-dashed lines represent CLOUDY predictions for
$Q$(H)=10$^{50.5}$~\smu~ and Kurucz (\cite{Kurucz94}) atmospheres with 35000 K
(lower curve) and 36000 K (upper curve).
Dashed lines are predictions for $Q$(H)=10$^{50.9}$~\smu~ and
{\it CoStar} atmospheres for 32600 K (lower curve) and  33300 K
(upper curve).
Dotted lines are the predictions for $Q$(H)=10$^{50.5}$~\smu~
with  {\it CoStar} atmospheres for 33300 K (lower curve) and  35500 K
(upper curve). }
\label{fig_Ndist}
\end{figure*}

Table \ref{tab_ratios}  lists the extinction corrected
\NIII\,57 to \NII\,122\,\mum~  and  \NeIII\,15.5 to \NeII\,12.8\,\mum~ 
ratios (hereafter \NIII/\NII~ and \NeIII/\NeII~  ratios, respectively).
The   extinction corrected  map of the
\NIII/\NII~ ratio  corresponding to the raster is
shown    in Fig.~\ref{fig_raster_coc}a.
The circles    in Fig.~\ref{fig_raster_coc}a indicate the location of 
the Quintuplet,  the  triangle indicates the location of the
Pistol Nebula and
the square  indicates the position of the Arches cluster.
The \NIII/\NII~ ratio ranges from 0.33 to 2.52.
For the range of \ne~ derived from the \OIII\,52/88 ratio,
the ratio of emissivities in
the two nitrogen lines ($\epsilon_{57}/\epsilon_{122}$) only
changes from $\sim 6.3$ to $\sim 7.7$ (Rubin et al. 1994).
Thus, in spite of the relatively low critical density of the  \NII~ line,
the \NIII/\NII~ ratio basically traces  changes in the
ionization  fraction    of \NIII~ relative to  \NII~ and not density variations.
Although the \NIII/\NII~ ratio shows a local 
maximum at the location of the Sickle, 
the absolute maximum is found towards the Pistol Nebula, close to
the Quintuplet (although it is elongated towards the southwest).
The \NIII/\NII~ ratio
 shows a general trend to decrease 
as the distance to the Quintuplet  cluster increases.
This is illustrated in Fig.\,\ref{fig_Ndist}, where we show all the
\NIII/\NII~ ratios as a function of their projected distance
to the Quintuplet cluster.
The \NIII/\NII~ ratio  shows a considerable dispersion which
could be due to the difference between projected and actual distances
and the possible minor  contribution of other ionizing  sources.
It is remarkable   that the degree of  ionization toward
sources like M1, M2, B4 and B5 located in the RAB,
to the south and east of the
Quintuplet, is similar to the ionization toward
sources located in other directions but at similar distances
from the cluster.
This clearly   points to the Quintuplet  as the main  ionization
source of the RAB and not only of the Sickle and the Pistol Nebula.

\begin{figure}[]
\resizebox{\hsize}{!}{\rotatebox{-90}{\includegraphics{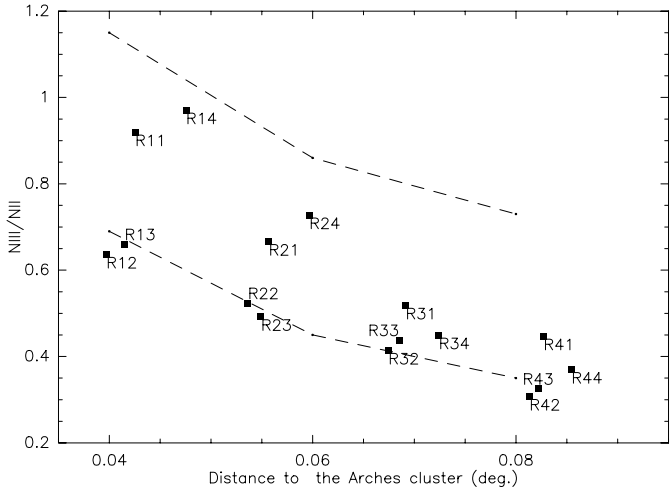}}}
\caption{  \NIII/\NII~  ratio derived from the first four columns of the
raster map  as a function of their projected distance to
the Arches cluster and comparison with models predictions
for using  $Q$(H)=10$^{51.4}$ \smu~  with  {\it CoStar}
atmospheres for 30400 K (lower curve) and  32600 K  (upper curve). }
\label{fig_NdistA}
\end{figure}

The northwest part of the \NIII/\NII~  map shows another 
gradient that  does not point to the Quintuplet but towards
other ionization source located  in the direction of the Arches cluster.
This cluster is believed to be the source of the ionization of the
Arched Filaments (Colgan et al. 1996; Cotera et al. \cite{Cotera00a}; Lang et al. \cite{Lang01}).
In fact, the western-most part  of the raster map covers the northern part
of the E1 arched filament (for notation see for instance Lang et al. \cite{Lang01}).
Figure\,\ref{fig_Ndist} also shows this effect. The \NIII/\NII~ ratio
measured for the raster points at distances larger than 0\fdg05 show
a systematic trend.
For a given column $y$ of the raster, the \NIII/\NII~ ratio  
decreases from the points in the 
first row of the raster (R1$y$)  to the points
in  the fourth column (R4$y$).
The effect is even more  clear when one plots 
the \NIII/\NII~ ratio derived 
for the first four columns of the raster as a function of their projected 
distance to the Arches cluster (see Fig.\,\ref{fig_NdistA}).

\begin{figure}[]
\resizebox{\hsize}{!}{\rotatebox{-90}{\includegraphics{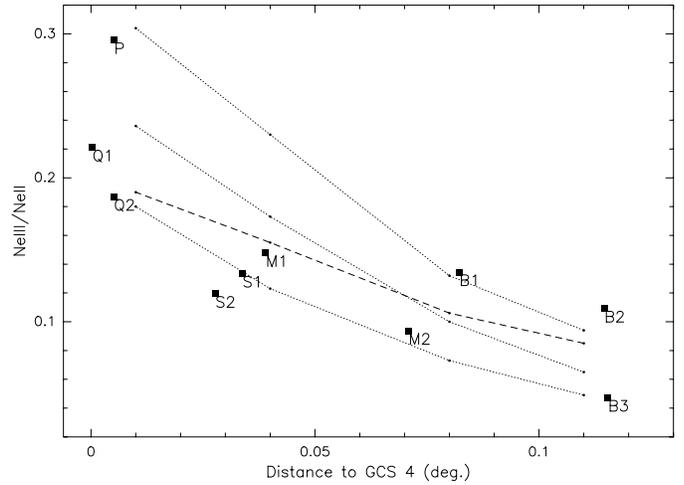}}}
\caption{NeIII/NeII ratio as a function of projected distance to
the Quintuplet source GCS4. Dotted lines are model predictions for
$Q$(H)=10$^{50.5}$ \smu~
with  {\it CoStar} atmospheres for 34500, 35000 and 35500 K
(from lower to upper curves).
Dashed lines are the predictions for  $Q$(H)=10$^{50.9}$ \smu~
with  {\it CoStar} atmospheres for 34500 K.}
\label{fig_Nedist}
\end{figure}

As for the \NIII/\NII~ ratio,
the maximum of the \NeIII/\NeII~  ratio is also found in
the Quintuplet and Pistol Nebula sources.
Figure~\ref{fig_Nedist} displays the \NeIII/\NeII~ ratio as
a function of projected  distance to the Quintuplet.
Although the number of available measurements
is less than those of the nitrogen lines,
the general behavior of the \NeIII/\NeII~  ratio is similar
to that of the \NIII/\NII~ ratio, decreasing as distance to the 
Quintuplet increases.
However, the gradient in the \NeIII/\NeII~ ratio with distance
is smaller   than that exhibited  by the \NIII/\NII~ ratio
($\sim 3$ instead of $\sim 10$).
The critical densities of both the \NeIII~ and the \NeII~ lines
are \gsim $2\,10^5$ \cmmt. Thus, for the range of \ne~ derived
for all the sources, the emissivities of  both lines do not depend on \ne,
and like the \NIII/\NII~ ratio, the \NeIII/\NeII ~ ratio traces
the ionization structure.
%
%


\section{Origin of the ionization}

To study the large scale ionization structure  
we have compared  the trend observed in the \NIII/\NII~  and \NeIII/\NeII~
ratios with the predictions  of photoionization models
considering the  Quintuplet and the Arches clusters as the main 
ionizing sources.
We have used CLOUDY version 90.04 (Ferland \cite{Ferland96})
and the  MICE interface developed by H. Spoon at the  MPE
to compute models for a  grid  of physical parameters.
Namely, effective temperatures,  \Teff, 
and   Lyman  continuum photons  fluxes, $Q$(H),
of the ionizing sources.
To define the shape of the incident continuum radiation
we have used the models of stellar atmospheres developed by Kurucz (1994)
and the Schaerer \& de Koter (\cite{Schaerer97}) 
``{\it CoStar}'' models, both  for   solar abundances.
We have   assumed  nebular N and Ne  abundances  of 3\,$10^{-4}$
and 2.8\,$10^{-4}$, respectively,  as derived by Rubin et al. (1995) for 
\HII~ regions in the GCR.
 These abundances are a factor of 2-3 higher than those measured
for Galactic disk \HII~ regions. We have checked with the 
Kurucz atmospheres  that increasing the stellar metallicities
in a factor of 3  changes the predicted lines ratios in less
than  15$\%$. Therefore, enhanced metal abundances in the
atmospheres does not affect significantly the results presented in
the following sections.
As a  typical density of the ionized material we have used
$10^{2.2}$\,\cmmt~  as  derived from the \OIII~ lines ratio.

\subsection{The Quintuplet cluster}

Since the Quintuplet cluster seems to dominate the ionization
for most of the observed positions, we first considered this 
cluster as the ionizing source.
We have used CLOUDY to estimate the size of the 
ionized region for the case
in which the ionizing source is surrounded by material with an uniform
\ne~ of 10$^{2.2}$\,\cmmt.
We assumed  $Q$(H)=10$^{50.9}$\,\smu~ as derived by Figer et al. (1999)
for the Quintuplet cluster 
and a \Teff~ of 36300 K (B2 model of Schaerer \& de Koter).
The predictions   of CLOUDY   for  the size
of the ionized region around the Quintuplet
is of \lsim 6 pc,   much smaller
 than what is observed from the fine structure line ratios  
(which is $\sim 18$ pc assuming a distance
to the Galactic center of 8.5 kpc).
One can, however, explain  the presence of \NeIII~ and \NIII~ 
at such   large distances from the Quintuplet  with the \Teff~ and  $Q$(H)
derived for the Quintuplet if  
the  material between the Quintuplet an the observed positions
has a density of  $\leq$ 10\,\cmmt.
This is consistent with  the  ring, or almost empty bubble,
morphology  observed  in the  MSX images.

We have therefore used a simple model
in which the ionized lines observed in our beam 
arise from  clouds of density 10$^{2.2}$\,\cmmt~ located at a distance
to the Quintuplet cluster equal to their projected 
distance on the sky (we assume a distance the GCR of 8.5 kpc).
Under these conditions,
one can characterize the effect of the  radiation
from a ionizing  source  on a cloud located at
a  distance $D$  using an ionization parameter ($U$) defined as: 
\begin{equation}
U=\frac{Q(\mathrm{H})}{4\pi D^2 n_e c}
\end{equation}
where $c$ is the velocity of light.

Following the models used by  Simpson et al. (1997) for  the Sickle
and the Pistol Nebula, we have first
computed the \NIII/\NII~ and \NeIII/\NeII~  ratios as a function of the 
distance of the clouds to the cluster (i.e. as a function of $U$)
for \Teff~ in the range  of 34000-38000 K
and $Q$(H) in the range of 10$^{49.5-50.5}$\,\smu~ 
with Kurucz (1994) atmospheres.
We found that $Q$(H)=10$^{49.5}$\,\smu~ produces too small \NIII/\NII~
ratios at   large distances.
On the opposite,  $Q$(H)=10$^{50.5}$\,\smu~   can explain the 
derived ratios if  \Teff~ is  in the range of 35000-36000 K (dot-dashed
lines in Fig.\,\ref{fig_Ndist}). 
These values are in good agreement with those of Simpson et al. (\cite{Simpson97})
and indicate that our results  extend to larger 
distances the results of Simpson et al.  derived for the Sickle
(see also Philipp et al.   in prep).
However, this model with the Kurucz atmospheres predicts
\NeIII/\NeII~  ratios which are a factor of $\sim 10$ smaller
than the measured values.
This is likely related to  the well known \NeIII/\NeII~  
problem associated to the Kurucz atmospheres,
which  predict \NeIII/\NeII~ ratios that  are always much smaller than the 
measured ones.
More recent model atmospheres  give larger \NeIII/\NeII~
ratios, much closer to the observed values (see e.g. Sellmaier et al. 1996). 

We have  tried  to 
explain both the \NeIII/\NeII~ and the \NIII/\NII~ ratios
consistently with similar  parameters for  the ionizing source 
using   CLOUDY models with $Q$(H)=10$^{50.9}$\,\smu~
using the stellar atmospheres of Schaerer \& de Koter.
Figure\,\ref{fig_Ndist}
shows, as dashed  lines, that the decrease of the \NIII/\NII~ ratio with
the distance is well explained with \Teff=32600-33300 K (B3 and A2 
{\it CoStar}  models).
The discrepancies between the observational results and the model
are within a factor of $\sim 1.5$.
This is quite reasonable in view of the simple model we use
and the uncertainties introduced by its approximations.
These are, mainly, the assumptions that
projected distances are    actual distances to the cluster and  
that there is not material in between the cluster and the observed sources
 (however, the model takes into account the attenuation of 
the radiation within the clouds).
In addition, the model uses  a single value of the  \ne~
for all the positions.
Figure\,\ref{fig_Nedist} shows, also  with dashed lines, that 
the  predictions of this model for the \NeIII/\NeII~ ratio
are in better agreement with the ratios derived from the 
observations than those derived using the Kurucz atmospheres.
However, the high   $Q$(H) value used  to explain the
\NIII/\NII~ ratios predicts  curves for  the 
\NeIII/\NeII~ ratio as a function of distance that  are rather flat.
Moreover, the  minimum \Teff~ of $\sim 34500$\,K  required
in order to explain the \NeIII/\NeII~ ratio 
is somewhat higher than the \Teff~ ($\sim 33000$ K) required to explain 
the \NIII/\NII~ ratio.

For completeness, Fig.\,\ref{fig_Ndist} and \ref{fig_Nedist} also show, 
as dotted  lines, the 
predictions for  model calculations using {\it CoStar} atmospheres
but with a slightly lower $Q$(H) of 10$^{50.5}$\,\smu.
Figure \ref{fig_Ndist} shows that we can reproduce the observed behavior
of the \NIII/\NII~ ratio for \Teff~ in the range of 33300 and 35500 K.
On the other hand, Fig.\,\ref{fig_Nedist} shows that we also  have a
rather good agreement between the predictions and the 
measured   \NeIII/\NeII~ ratio for  \Teff$\sim 34500-35500$ K,
which is similar to the \Teff~ needed to explain the \NIII/\NII~ ratio.
In summary, the \NIII/\NII~ and \NeIII/\NeII~ ratios observed with ISO
in the RAB region 
can be explained by assuming that the fine structure lines arise
from clouds located at different distances from the Quintuplet cluster,
which would be the main ionizing source.
The required Lyman continuum photons flux of $10^{50.5-50.9}$ \smu~
and effective temperatures of $\sim 35000$\,K are in good
agreement with the estimations based on the stellar content of the
cluster (see Figer et al. 1999).

\subsection{The Arches cluster}

As mentioned in section 4.3 the observed gradient  in the
\NIII/\NII~ ratio  in the northwest region of the raster
cannot be explained by the effect of  the Quintuplet.
In this section we will discuss the influence
of the Arches cluster in the ionization of the Radio Arc material.
Figure~ \ref{fig_NdistA} shows 
the \NIII/\NII~ ratio derived for the positions of the
raster map which are closer to the Arches cluster 
as a function of their projected distance to this cluster.
Figure\,\ref{fig_NdistA} also displays, as long-dashed lines,
the results of  CLOUDY
models computed with {\it CoStar} atmospheres, 
$Q$(H)=10$^{51.4}$ \smu~  and \ne=10$^{2.2}$ \cmmt.
The observed gradient in the \NIII/\NII~ ratio  is
pretty well explained with a \Teff~ in the range of 30400-32600 K.
These parameters are in agreement with those estimated for
the Arches cluster  (Morris et al. in prep see also Lang et al. 2001).
Therefore, we conclude  that  the observed 
\NIII/\NII~ gradient to the southwest 
is likely produced by the Arches  cluster and 
that the influence   of the radiation from this       cluster should
be considered up to distances of 7-13 pc.

\subsection{The combined effect of both clusters}

We now consider the combined effects of  the two clusters on  the 
ionization of the whole region surrounding them.
As discussed in Sect. 5.1, the \NIII/\NII~ and \NeIII/\NeII~ ratios
are directly related to the ionization parameter $U$.
Since the effective temperature of the radiation derived for
both clusters are rather similar (see above), we have considered
the most  simple model to account for the combined effect
of the two clusters. 

As a first approximation one can estimate a ``total  
ionization parameter" ($U_\mathrm{tot}$) as the sum of
two different ionization parameters, one due to the
Quintuplet ($U_\mathrm{Q}$) and other due to the Arches cluster
($U_\mathrm{A}$).
Thus, for a cloud     
at distances  $D_\mathrm{Q}$ and $D_\mathrm{A}$ from the
Quintuplet and the Arches cluster, respectively, $U_\mathrm{tot}$
will be given by:
\begin{equation}
U_\mathrm{tot}=U_\mathrm{Q}+U_\mathrm{A}=
   \frac{Q(\mathrm{H})_\mathrm{Q}}{4\pi D^2_\mathrm{Q} n_e c} + 
   \frac{Q(\mathrm{H})_\mathrm{A}}{4\pi D^2_\mathrm{A} n_e c}
\end{equation}   
We have plotted in the upper panel of  Fig.\,\ref{fig_UX} some  contours of 
equal $U_\mathrm{tot}$ assuming a constant density of  \ne=10$^{2.2}$ \cmmt, 
a $Q(\mathrm{H})_\mathrm{Q}$ of $10^{50.9}$ \smu~ 
and a  $Q(\mathrm{H})_\mathrm{A}$ of $10^{51.4}$ \smu.  
The agreement of the iso-$U_\mathrm{tot}$  curves  with the contour map 
of the \NIII/\NII~ ratio (shown with thick lines)
is very  good taking into account the simplicity of the model,
which for instance, does not consider any shielding of the radiation. 
It reproduces the observed gradients towards the Arches and Quintuplet
clusters.
However, the model does not account for the local maxima found in
the Sickle.
The lower \NIII/\NII~ ratios derived towards points located southward 
(in galactic latitude) of 
the Quintuplet which lie on the same iso-$U_\mathrm{tot}$
curve than others located northward
of the cluster should be due to the fact that $U_\mathrm{tot}$ is more
homogeneous in the LWS beam 
in the  north of the cluster than in the  south.

Furthermore, this simple model also reproduces  the observed overall
distribution of the warm dust.
This is also  illustrated in the upper panel of 
Fig. ~\ref{fig_UX}, where the iso-$U_\mathrm{tot}$
contours are  superposed on 
the 25\,\mum~ band image taken by MSX.
One can see    that the morphology of
the warm dust emission approximately follows the shape of the
iso-$U_\mathrm{tot}$ curves.
The warm dust in the area of the Thermal Filaments
is under the strong  influence of the Arches cluster while the warm dust
emission in the Sickle follows the shape of the iso-$U_\mathrm{tot}$
curves that are dominated by the Quintuplet.
The elliptical rather than circular distribution of the warm dust
around the clusters, in particular around the Arches cluster,
can be explained by the combined effect of both clusters.
Even at large distances from the clusters (\gsim  15 pc)
the distribution of the warm dust also  seems to follow
the shape of the iso-$U_\mathrm{tot}$
curves, see for instance in Fig. \,\ref{fig_UX}
the  25\,\mum~ dust emission in
$(l,b)\sim (0\fdg2, -0\fdg15)$ or in $(l,b)\sim (0\fdg05, -0\fdg12)$.
The agreement between  the morphology of the warm dust emission with 
the iso-$U_\mathrm{tot}$ curves strongly suggest a  same origin
for both the warm dust and the ionized gas (see next section).

\begin{figure*}
\vspace{15cm}
\caption{{\it Upper panel:} Contours of equal total ionization parameter
(iso-$U_\mathrm{tot}$ curves, see Sects. 5 and 6)  
 overlaid on the 18-25 \mum~ band image of MSX.
The display is logarithmic from $10^{-6}$ to 2 $10^{-3}$ W\,m$^{-2}$\,sr$^{-1}$.
The contour levels ($\log U_\mathrm{tot}$) are: -0.5, -1, -1.15, 
-1.22, -1.3, and from  -1.4 to -2.15 by -0.15).
Thick contours are the \NIII/\NII~ map as represented in
Fig.\,\ref{fig_raster_coc}.
Note that the model reproduces the observed gradients in the
\NIII/\NII~ ratio toward the Quintuplet and the Arches cluster 
(both represented as filled squares).
In addition, the warm dust emission seems to follows the morphology
of the iso-$U_\mathrm{tot}$ curves.
{\it Lower panel:} Fe 6.4 keV line emission map of Koyama et al. (1996)
overlaid in the 18-25 \mum~ band  MSX image.
The contour levels of the Fe 6.4 keV line map in
units of $10^{-6}$ counts/sec/0.106\,min$^2$ are: 0.25 and from 0.4 to
1.1 by steps of 0.1).
Note that the Fe 6.4 keV  emission  around
$(l,b)\sim(0\fdg13,-0\fdg11)$ fills the bubble.
The positions of 1E 1742.9--2849 and 1E 1743.1--2852 are shown by
 a solid circle  and a triangle, respectively.}
\label{fig_UX}
\end{figure*}
%


\section{The Radio Arc bubble of warm dust}

The origin of the ring or bubble  of warm dust it is not clear
but the ISO data presented in this paper 
shows that the ionization of the gas
is mainly produced by the Quintuplet instead of 
any source in the center of the ring.
One unlikely possibility is that the hot dust observed in the 
RAB is not heated by the UV radiation from the clusters
that explain the ionization structure (see Sect. 5.3).
In this case, one would expect not to see a  correlation between
the warm dust and the ionization. 
We can check this possibility by comparing for instance the 
\OIII\,88\,\mum~ line intensity with the continuum emission from 
warm dust for different positions in the RAB.
The intensity of this fine structure line
for points ``inside the ring" (M1, R19, R29; i.e.  where
the emission from warm dust is relatively weak, 
see Fig\,\ref{fig_obs})
is lower than  that for points like M2, B4 or B5  
located on the edge of the ring .
For instance, the average  \OIII\,88\,\mum~  line fluxes for  the
points on the ring is $\sim 12\,10^{-18}$\,W\,\cmmd~ 
while for those  points inside the ring is $\sim  7\,10^{-18}$\,W\,\cmmd.
From the IRAS HIRES 25 \mum~ map (with a resolution of 
$66^{''}\,\times\,35^{''}$)
one obtains  that the averaged intensity of the 
dust emission at this wavelength for the points on the edge of 
the ring,  M2, B4 and B5, 
is $\sim 1500$ MJy\,sr$^{-1}$ and $\sim 750$ MJy\,sr$^{-1}$
for the points inside the ring (M1, R19, R29).
Furthermore, towards the  Sickle positions (R16-R46)
the average  \OIII\,88\,\mum~  line flux 
is $\sim 36\,10^{-18}$\,W\,\cmmd~ while the 25 \mum~ intensity
is $\sim 5000$ MJy\,sr$^{-1}$.
Therefore, it seems that the ratio of the luminosity of the fine structure
lines (ionized gas column density)
 to the warm dust emission  is approximately constant
for all positions in the RAB area.
This constant line-to-continuum ratio shows that the warm dust
should  also be heated by the UV radiation from the cluster.
The apparently lack of warm
dust in the central part of the ring is just an effect of
column density but not of UV radiation since the \NIII/\NII~
ratio is higher for points like M1 inside the RAB, 
than for points at the edge of the RAB like M2.

The smaller column density of material towards the  center
of the ring is consistent with different geometries like a ring 
(or a cylinder with its axe close to the line of sight) and 
a shell or almost  empty bubble.
For  a  shell geometry  with constant density and  internal and 
external radius of 0\fdg05 and 0\fdg075, respectively, 
the expected  column density 
in the line of sight to the edge of the shell is
a factor of $\sim 2$  larger
than that  in the line of sight towards the center of the shell.
This is  the typical ratio of the \OIII\,88\,\mum~ line or the 
25 \mum~ dust emission between M2, B4, B5 at the edge of the
shell  and M1, R19, R29 at the center.
Although this does not demonstrate that the observed structure
has a shell morphology  instead of a ring one,
the data are  consistent with a shell morphology (bubble)
of dust heated, and gas ionized, by the UV radiation of the
Quintuplet and the Arches cluster.

The geometrical  center of the  bubble is approximately located
at $(l,b)=(0\fdg16,-0\fdg11)$, over the NTFs.
We have searched in the SIMBAD database for sources in the central
area of the RAB  without success. 
The closest sources to the geometrical  center of the 
bubble  are the near-IR \object{ sources
48 and 49 of Nagata et al. (\cite{Nagata93})} and our source M1
but no link between them and the RAB    can be
established. 
Somewhat further from the center there are other IR sources
detected by  Nagata et al. (1993) like \object{Objects 43-46} 
but they are late type M-stars candidates.
However, close to this sources,
there is  also a luminous X-ray source:
\object{1E\,1742.9-2849} (represented as a filled circle in the
lower panel of Fig.~\ref{fig_UX}), with an X-ray luminosity 
of $\sim 8\,10^{34}$  erg\,\smu~ (Predehl \& Tr\"umper \cite{Predehl94},
also recently detected by {\it Chandra}, Yusef-Zadeh priv. communication).
It is $\sim 0\fdg04$ ($\sim 6$~ pc) far from the center of the shell and
its position  suggests that it could  be responsible 
for the elongation of the maximum in the \NIII/\NII~ ratio
toward the southwest
instead of being centered in the Quintuplet  and the Pistol Nebula
as expected from the models.
There is another X-ray source inside the shell at a similar distance
from its center: \object{1E\,1743.1-2852} (triangle in the lower panel of
Fig.~\ref{fig_UX}).

It is remarkable that the diffuse hard X-rays (2-10 keV)
continuum emission observed by the ASCA satellite 
towards the GCR (Koyama  et al. \cite{Koyama96})
shows extended emission which covers  an important fraction of the RAB.
This would indicate that the RAB might be filled with hot gas that gives rise to
the extended X-ray emission.
One  interesting prediction for the proposed scenario
of a bubble filled with hot gas
is that  the morphology of the 6.4 keV line of neutral Fe,
which traces gas  irradiated by hard X-rays 
(that is {\it X-ray dominated regions} -XDRs)
should trace the walls of the bubble.
The lower panel of Fig.~\ref{fig_UX} shows 
the Fe 6.4 keV line emission map of Koyama et al. (1996)
overlaid on the MSX image.
The  Fe 6.4 keV line emission  shows two maxima in the Sgr A
neighborhood. One is located at $(l,b)\sim (0\fdg04, -0\fdg01)$,
contouring the hard
X-rays emission of the SgrA region while the other maximum is located
at $(l,b)\sim (0\fdg13,-0\fdg11)$ within the RAB.      
The Fe 6.4 keV emission in the RAB 
peaks in between 1E\,1742.9-2849 and
1E\,1743.1-2852, although  closer to the last source.
It clearly fills  the bubble covering
the region between the NTFs at $l\sim 0\fdg18$
and the edge of the bubble at $l\sim 0\fdg08$.

In summary, the RAB seems to be  filled with  hot gas emitting hard 
X-rays continuum which create a large XDR that is revealed 
by the Fe 6.4 keV line emission.
We speculate with the possibility that the X-ray sources 
might be related with the  energetic events that created the bubbles
of warm dust.
Relative motions of the ISM and the X-ray source would explain 
the off-center position of the X-ray sources in the RAB.
Considering the shape, the size  of the bubble and the presence
of the X-rays Fe lines,
a supernova explosion is the most likely origin for the RAB.


\section{General implications}

As suggested by Mart\'{\i}n-Pintado et al. 2000a, 
the picture arising from the comparison of the observed
ionization and the model calculations is that the
ISM in the Radio Arc region is highly inhomogeneous
as indicated by 
the presence of bubbles filled with relatively
low density ionized gas and hot gas emitting in X-rays.
This structure allows the radiation from hot stars
to reach large distances giving rise to extended low density 
ionized regions.
The Quintuplet and the Arches cluster
are the dominant sources of  ionization in the Radio Arc region.
Alternative ionization mechanisms
(for instance magnetic effects) would play a minor role and
should be restricted to smaller regions such as the Sickle.

This inhomogeneous structure seems to be a typical
characteristic of the GCR ISM.
In fact, this kind of structures  has been invoked to explain the
non-detection of hydrogen recombination lines and the detection
of fine-structure  lines  from ions
like \NeII, \SIII~ or \OIII~ in a sample of clouds distributed
all along the GCR (Mart\'{\i}n-Pintado et al. \cite{MP00a};
Rodr\'{\i}guez-Fern\'andez et al.  in prep.).
On the other hand, inhomogeneities like
shells and filaments are known to exist in other regions of the
GCR like those revealed by the CO survey of
Oka et al. (1998)
or those found by Mart\'{\i}n-Pintado et al. (\cite{MP99}) in the
envelope of Sgr B2 which,  like for the RAB,  also seem to be correlated with 
the Fe 6.4 keV line emission (see  Mart\'{\i}n-Pintado et al. \cite{MP00b}).

\section{Conclusions}

We have presented ISO observations of selected fine structure lines in the
Radio Arc in the Galactic center region.
The analyzed lines are \NIII~57\,\mum, \NII~122\,\mum, \OIII~ 52 and
88\,\mum, \NeIII~15.6\,\mum, \NeII~12.8\,\mum~ and \SIII~ 18.7 and 33.5\,\mum.
The main results derived from these observations can be summarized as
follows:

\begin{itemize}
\item The electronic densities derived from the \OIII~ 52 to 88\,\mum~ ratio
range between $10^{1.8}$ and $10^{2.6}$ \cmmt.
The electronic densities derived from the \SIII~ 18.7 to 33.5 \mum~ ratio
in the smaller beam of the SWS
are $\sim 10^2$ \cmmt~ for all the sources but those in the Quintuplet
and Pistol Nebula which reach values of  $\sim 10^{3.5}$ \cmmt.

\item After correcting for extinction, the \NIII/\NII~ and the
\NeIII/\NeII~ ratios
vary from position to position from 0.3 to 2.5  and from 0.05 to 0.30,
respectively.
Both ratios exhibit a clear trend to decrease with increasing distance
to the Quintuplet cluster.
For some observed positions,
the \NIII/\NII~ ratio also shows a gradient that points towards
the Arches cluster.

\item Photoionization models confirm that the observed ratios and their
trend with the distance to the clusters
are consistent with the derived  parameters
for  the clusters, that is \Teff=32000--35000~K, and a flux of Lyman
continuum photons of $\sim 10^{51.4}$~\smu~ and $\sim 10^{50.9}$~\smu~ for
the Arches and the Quintuplet cluster, respectively.
Thus, the combined effects of both the
Quintuplet and the Arches cluster
completely dominate  the ionization in the Radio Arc region.
They create a large ionized region with a size of more than 
$30\times30$~pc$^2$.
Alternative ionization mechanisms
(for instance magnetic effects or more hot stars) seems to
play a minor role at large scales.

\item A simple model of a total ionization parameter
expected for  the two clusters
can explain not only the ionization structure but also
the distribution of warm dust and, in particular, the elliptical
rather than circular symmetry of some features like the Thermal
Filaments.
The warm dust seems to be correlated with the ionized gas and therefore
also seems to be heated by the  two clusters that ionize the gas.

\item The effect of the radiation from the  Quintuplet stars  over large
distances is at least in part due
to the presence of a shell or an  almost empty bubble
(Radio Arc Bubble, RAB) of warm dust and ionized gas.

\item The RAB is filled with hot gas emitting hard X-rays which
creates a large X-ray dominated region (XDR) observed in the
Fe 6.4 keV line emission.

\item The origin of the bubble it is not clear,  but it could be
related to  a couple X-ray sources located  $\sim 6$~pc
off the shell center.
Taking into account the size and shape of the bubble and the presence
of the Fe X-rays lines, a supernova origin is certainly possible.

\end{itemize}

\begin{acknowledgements}

   NJR-F acknowledges  {\it Consejer\'{\i}a de
Educaci\'on  de la  Comunidad de Madrid} for a predoctoral fellowship.
   The authors acknowledge support  by the {\it Ministerio de 
Ciencia y Tecnolog\'{\i}a}
under grants  1FD97-1442 and PNE 014-2000-C.
   NJR-F thanks H. Spoon for his introduction to CLOUDY and MICE.
   MICE, SWS and the ISO Spectrometer Data Center at MPE are
supported by DLR (DARA) under grants 50 QI 86108 and 50 QI 94023.
   The ISO Spectral Analysis Package (ISAP) is a joint development by the
LWS and SWS Instrument Teams and Data Centers. Contributing institutes
are CESR, IAS, IPAC, MPE, RAL and SRON.
   This research has made use of the SIMBAD database,
operated at CDS, Strasbourg, France.
   This research made use of data products from the Midcourse Space Experiment.
Processing of the data was funded by the Ballistic Missile Defense
Organization with additional support from NASA Office of Space 
Science.  The data were accessed by
services provided by the NASA/IPAC Infrared Science Archive.
\end{acknowledgements}


\end{document}